\documentclass{jaa}
\usepackage{natbib}
\usepackage{placeins}
\usepackage{subcaption}
\usepackage{gensymb}
\usepackage{multirow}
\usepackage{hhline}
\usepackage{parskip}
\usepackage{xcolor}
\usepackage{siunitx}
\usepackage{float}

\usepackage{graphicx}
\usepackage{mwe} 
\usepackage{comment}

\usepackage{textcomp}

\usepackage{hyperref}
\hypersetup{
    colorlinks=true,
    linkcolor=blue,
    filecolor=magenta,      
    urlcolor=cyan,
    pdftitle={Overleaf Example},
    citecolor=red,
    }
\usepackage{graphicx}

\begin{document}\sloppy

\title{Simple Two-Element LPDA Radio Interferometer at Gauribidanur Observatory: An Initial Results} 

\title{Initial Results from the Two-Element Interferometer at Gauribidanur Radio Observatory} 

\title{Solar Bursts: A Simple Two-Element Interferometer at Gauribidanur Observatory using RRI LPDA} 

\title{A Simple Two-Element Radio Interferometer with LPDAs at the Gauribidanur Observatory and Solar Observation Results} 

\title{A Simple Two-Element Radio Interferometer at RRI Gauribidanur Observatory with Solar Event observation} 

\title{GBD2E-1: A new 180 to 350 MHz Two-Element Radio Interferometer Observing Solar Events at Gauribidanur} 

\title{TIGRS: Two-element Interferometers at the Gauribidanur Radio Station  }

\title{TWIN-1: A new 185 to 350 MHz Two-Element Radio Interferometer for Solar Observation at Gauribidanur}


\author{
Abhishek R\textsuperscript{1},
G N Paneendra\textsuperscript{1,2},
Raghavendra Rao K B\textsuperscript{1},
Keerthipriya S\textsuperscript{1}
Katta Naveen\textsuperscript{1,3},
Anumanchi Agastya Sai Ram Likhit\textsuperscript{1,3},
Arul Pandian B\textsuperscript{1},
Vinod K R\textsuperscript{1},
Roshan Kumar Dora\textsuperscript{1,3},
Arutkeerthi N\textsuperscript{1,4},
Deep Bhowmik\textsuperscript{1,5},
Krishna Balaji S\textsuperscript{1,6}, 
Sahana Bhattramakki\textsuperscript{1}, 
Prabu Thiagaraj\textsuperscript{1}}
 
 \affilOne{\textsuperscript{1} Electronics Engineering Group, Raman Research Institute, 560080, India \\}

 \affilTwo{\textsuperscript{2} Department of Physics and Electronics, Christ (Deemed to be University), Bangalore 560029, India\\} 
 
\affilThree{\textsuperscript{3}  Department of Physics, Indian Institute of Science Education and Research Bhopal, 462066, India\\}

\affilFour{\textsuperscript{4} Department of Physics, Indian Institute of Technology Roorkee, Roorkee 247667, India\\} 

\affilFive{\textsuperscript{5} Department of Physics, Indian Institute of Science, Bangalore 560012, India\\}  
 
\affilSix{\textsuperscript{6} Department of Physics, Indian Institute of Technology Madras, Chennai 600036, India\\}


\twocolumn[{

\maketitle

\corres{abhishek@rri.res.in}

\begin{abstract}

We present details of a new two-element radio interferometer, namely the TWIN, developed for routine monitoring of solar events at the Gauribidanur Observatory (77.428°E, 13.603°N) between 185 MHz and 350 MHz. This interferometer is built using custom-built antennas, an indigenous RF front-end consisting of LNAs, power combiners, and high-gain amplifiers, and RF over fiber signal transport modules. A data capture system was developed using a commercial digital storage oscilloscope and controlled with custom-designed Python-based tools.    The oscilloscope digitises the RF signals with a time resolution of 0.8 ns, bursts of 9952 time samples, with a nominal duty cycle of 3 seconds, and the resultant data is collected on a remote computer. Specialised tools were developed to analyse the data and detect solar events. The interferometer has been operational since July 2023. In our analysis, we have detected several solar events from the ongoing Solar Cycle 25. This paper outlines the design philosophy, signal flow, commissioning and validation, results from a solar event, and future plans for data capture, data analysis, and archival.

\end{abstract}

\keywords{Antenna Design;  Log-Periodic Dipole Antennas;  Radio Astronomy Observations; Radio Frequency Interference (RFI); Radio Interferometry; Solar events. }

}]


\doinum{12.3456/s78910-011-012-3}
\artcitid{\#\#\#\#}
\volnum{000}
\year{0000}
\pgrange{1--}
\setcounter{page}{1}
\lp{1}

\section{Introduction}

Low-frequency radio astronomy is growing in popularity and plays a critical role in the study of astronomical sources. A few large radio telescopes (LOFAR, MWA and LWA) around the world operate at low frequencies below 350 MHz. A simple 20 MHz two-element interferometer, SIMPLETON, was operational in the past \color{red}\citep{simpleton}\color{black}.  We present the development and commissioning of a simple low-frequency two-element Log-Periodic Dipole Array (LPDA) interferometer, TWIN, at the  Gauribidanur Radio Observatory (77.428° E, 13.603° N) located approximately 85 km north of Bangalore, with an elevation of  694 m above sea level. The GeeTee, GRAPH, and DART \citep{DART} are among the radio telescopes at the observatory. The GRAPH is an imaging telescope routinely used to observe the Sun. While GeeTee was originally an imaging instrument, was later converted into a time-domain instrument, and ceased operations after Covid-19, the DART is operational for pulsar studies.  The TWIN is a newly constructed two-element interferometer operating between 185 MHz and 350 MHz, commissioned at the Gauribidanur observatory, with student participation. It integrates custom-built LPDAs, an in-house developed RF front-end receiver system, and an oscilloscope-based data acquisition system. A Python-based pipeline enables automated observation scheduling, real-time monitoring, and preliminary data analysis \citep{likhit2025innovativewebtoolremote}. Over the last two years of operation, the interferometer has demonstrated robust performance, observing solar transits and detecting numerous solar events.

The Sun’s activity fluctuates in cycles of 11 years. Recently, there has been an increase in the Sun's activity during the present Solar Cycle 25. As a result, we are observing a higher number of events, such as solar flares, solar storms, and coronal mass ejections, that influence near-Earth space weather \citep{Jasinski_2025}. Given these changes, observations during these periods are crucial for understanding their impacts and analysing the underlying physical processes. The TWIN interferometer routinely monitors the Sun as it transits across its zenith.

This paper outlines the design, implementation, and initial results from commissioning the interferometer at Gauribidanur Radio Observatory. Section 2 provides an introduction to the interferometer. Section 3 outlines the interferometer signal chain, consisting of an LPDA, an LNA, a front-end RF receiver, an RF-over-Fibre (RFoF) transceiver module, and the data capture system. Section 4 outlines the performance and characterisation of the hardware involved, showing the RFI situation at the Gauribidanur Radio Observatory. Section 5 presents the interferometric observations and highlights salient findings; Section 6 outlines future directions, followed by the conclusion in Section 7.

\section{Introduction to Interferometer}
An interferometer consists of arrays of antennas to produce higher-resolution images. In an interferometer, the far-field signals arrive at the antennas as plane-wave fronts, reaching different antennas at different times depending on the source direction and the antennas' orientation. As a result, the signals from the different antennas acquire phase differences that must be compensated before combining them to form a phased-array output or correlating them for an interferometer output.   Hence, in an interferometer antenna setup, suitable delays, known as tracking delays, are introduced in the signal path, thereby steering the phase centre of the antennas to track the direction of the source of interest.  Such tracking of the signal path is essential for achieving the maximum response of the combined and correlated outputs from the antennas .   The TWIN interferometer records the voltages from the two elements separately, allowing such delay tracking to be applied offline during post-processing, if needed. A MATLAB-based model for calculating the tracking delays in an interferometer is developed \citep{SB1}.

\section{Interferometer Signal Flow }

\begin{figure*}[h!]
    \centering
    \includegraphics[width=\linewidth]{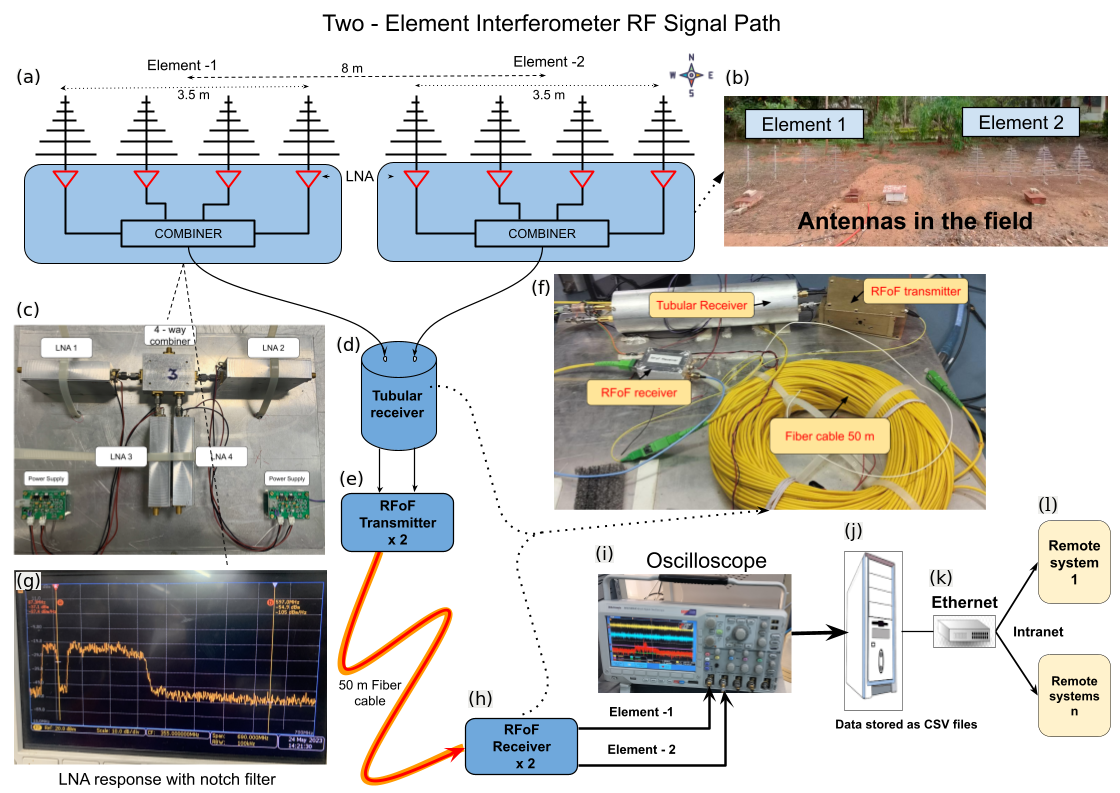}
    \caption{Block diagram of the TWIN interferometer RF signal path from the Antennas to the Digitization. (a) Schematic representation of 8 Log Periodic Dipole Antennas (LPDAs) arranged as two elements, each element containing 4 LPDAs. (b) Real-life deployment of the 8 LPDAs. (c) A four-way combiner. (d) Tubular receiver. (e) Two RFoF transmitters convert analogue RF signals to digital optical signals. (f) Tubular receiver, the RFoF transmitter, receiver and optical fiber as deployed. (g) Response of the LNA showcasing the notch filter it contains. (h) The RFoF receiver stage. (i) Oscilloscope, which is used for digitisation and data writing. (j) The data is stored in a computer, and it can be accessed through (k) Ethernet. (l) Other computers present in the intranet network can access the data.}
    
    \label{fig:BLOCK_DIAG}
\end{figure*}

The TWIN interferometer consists of two elements, each made by combining four LPDAs as shown in Figure \ref{fig:BLOCK_DIAG}a,b.  The interferometer is installed along an east–west baseline to monitor transiting celestial sources. 

\begin{figure*}[h]
    \centering
    \includegraphics[width=18cm, height =11 cm]{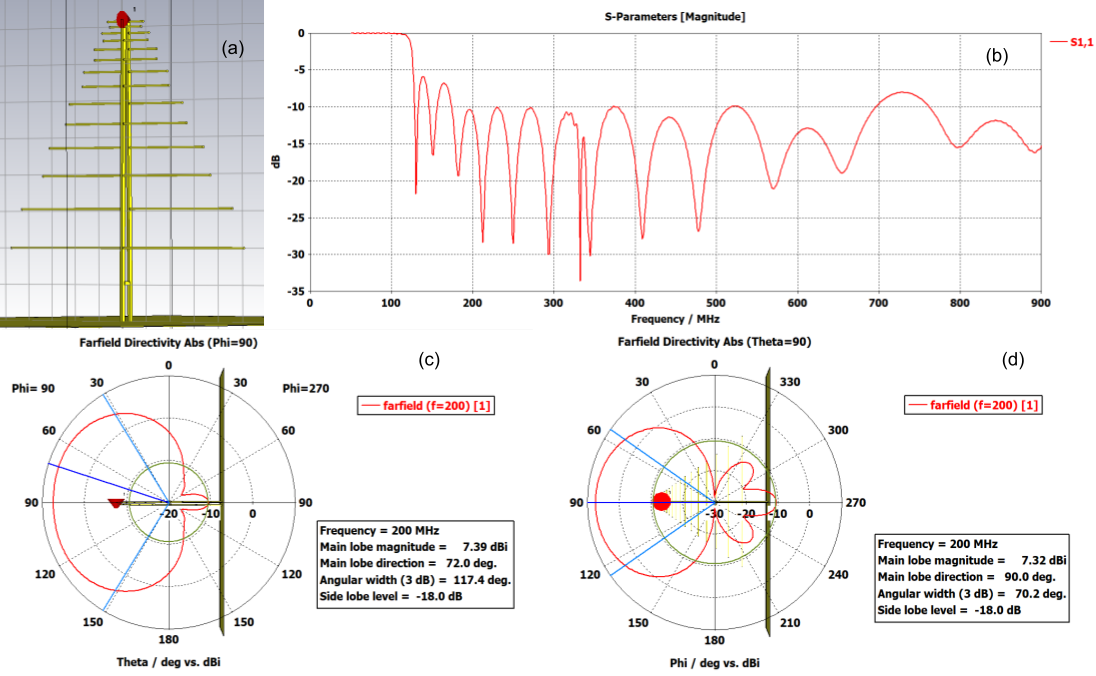}
    \caption{ (a) The log periodic dipole antenna. (b) S\textsubscript{11} response of the antenna. (c) H-Plane beam response at 200 MHz. (d) E-Plane beam response at 200 MHz. }
    \label{fig:single_antenna}
\end{figure*}

The LPDA, as shown in Figure \ref{fig:single_antenna}, has 11 dipoles designed for operation in the 185–350 MHz band. The dipoles are 9 mm-diameter aluminium tubes; the booms are 11 mm-square aluminium tubes, each 840 mm long. The LPDA boom separation is 21 mm at the low-frequency end and 6 mm at the high-frequency end, and maintains a 50-ohm characteristic impedance at the feed. Each LPDA in a group is spaced 1.08 m apart, and the two elements (groups of four LPDAs) of the interferometer are separated by 8 m along the East-West direction. 

The celestial signals received by the LPDAs are inherently weak, requiring amplification and filtering to suppress interference and enable transmission to the lab for recording and subsequent data processing.  The RF signals from the LPDAs are first amplified by low-noise amplifiers (LNAs). The LNA output signals from four LPDAs (corresponding to each element) are summed using a 4-way RF combiner. The element outputs thus obtained are further amplified independently in the high-gain front-end RF receiver, housed inside a tubular module and transmitted to the lab using a custom-developed RF over Fiber transceiver.  At the lab, a digital storage oscilloscope is used for digitisation, and a computer is used to record the data. An outline of the interferometer signal flow is shown in Figure \ref{fig:BLOCK_DIAG}. The front-end RF signal chain for the interferometer, consisting of the low-noise amplifiers (LNAs), signal combiners,  high-gain front-end RF receiver, and the RF over Fibre transmitters, is developed in-house.  The LNAs have a gain of 11 dB, a noise figure of 0.6 dB, and filters to suppress signals below 45 MHz and between 88 and 108 MHz (the FM band), thus preventing strong RFI from affecting the signal chain. The high-gain amplifiers used in the interferometer signal chain have a gain of 53 dB and incorporate an additional bandpass filter to restrict the observation band to 185-350 MHz and to avoid communication RFI outside this range.

An RFoF transmitter module employs a pigtailed laser diode (1310 nm, 4 mW output power) to convert the electrical RF signal into an optical signal.  After transmission over 50 m on a single-mode fiber cable from the field to the lab, the optical signals are converted back to electrical signals using the RFoF receiver module for digitisation by the multi-channel digital oscilloscope data-acquisition system. The overall gain of the RFoF transceiver is around 13 dB.

An oscilloscope synchronously digitises the two-element signals at a 1.25 GHz sampling rate and, in burst mode, transfers blocks of 8-bit-wide, 9,952-sample voltage samples from each element once every 3 seconds to a remote computer. The digitised signals are monitored in real time and stored as packets for further analysis.

Interferometer data collection is based on a Python-based data acquisition suite that enables automated observational scheduling, real-time monitoring, and preliminary data processing. This system provides control of the data acquisition via an SCPI command sequence using the Python libraries Streamlit and PyVISA. This interface also allows remote access via a secure WLAN network, supports user-initiated observations, and provides features such as manual RFI masking, transit plotting, and fringe-plot analysis \citep{likhit2025innovativewebtoolremote}.

\section{Performance and Characterization}

Single LPDA and the individual element beam patterns were simulated in CST-Studio\textsuperscript{\textregistered} and the interferometer responses were studied using a MATLAB model \citep{Matmodel}. The front-end electronics and the fiber transceiver modules were developed in-house.

\subsection{Simulated response of the LPDA using CST }

 The main-lobe gain, half-power beam width and primary side-lobe level, for H and E-planes obtained at 150, 200, 250, 300, and 350 MHz, corrsponding to individual LPDA in Tables \ref{tab:my_table2} and \ref{tab:my_table1} respectively, and for and element (groups of four LPDAs) in Tables \ref{tab:my_table4} and  \ref{tab:my_table3}. Figure \ref{fig:single_antenna} b shows the S\textsubscript{11} response for a single LPDA, Figures\ref{fig:single_antenna} c and d show the LPDA's H- and E-plane responses at 200 MHz.

    \begin{table}[!htbp]
    \centering
    \renewcommand{\arraystretch}{1.3}
    \title{\textbf{Single LPDA Response H-Plane}}\\
    \vspace{4 mm}
    \resizebox{\columnwidth}{!}{%
    \begin{tabular}{|c|c|c|c|c|c|}
    \hline
    \multirow{2}{*}{\textbf{Parameter(Units)}} &\multicolumn{5}{c|}{\textbf{Frequency in MHz}}
    \\
    \hhline{~-----}
    & \textbf{150} & \textbf{200} & \textbf{250} & \textbf{300} & \textbf{350 }\\
         \hline 
      Main Lobe(dBi) & 8.7 &7.4 & 6.4& 7.9 & 7.0\\
        Side Lobe(dB)& -19.2 &-18&-18.1& -2.2 & -2\\
        Half-power beam width & 84.5 & 117.4&147.4 & 57.6&88.5 \\
         \hline
    \end{tabular}}
    \caption{The H-plane main-lobe, side-lobe and half-power beam widths obtained for an LPDA placed over a ground plane.}
    \label{tab:my_table2}
    \end{table}

        \begin{table}[!htbp]
    \centering
    \renewcommand{\arraystretch}{1.3}
    \title{\textbf{Single LPDA Response E-Plane}}\\
    \vspace{4 mm}
    \resizebox{\columnwidth}{!}{%
    \begin{tabular}{|c|c|c|c|c|c|}
    \hline
    \multirow{2}{*}{\textbf{Parameter(Units)}} &\multicolumn{5}{c|}{\textbf{Frequency in MHz}}
    \\
    \hhline{~-----}
    & \textbf{150} & \textbf{200} & \textbf{250} & \textbf{300} & \textbf{350 }\\
         \hline 
      Main Lobe(dBi) & 8.7 &7.3 & 4.3& 7.9 & 6.8\\
        Side Lobe(dB)& -19.2 &-18&-16& -19.1& -17.3\\
        Half power beam width & 58.8 & 70.2&104.4 & 42.8&72.7 \\
         \hline
    \end{tabular}}
    \caption{The E-plane main-lobe, side-lobe and half-power beam widths obtained for an LPDA placed over a ground plane.}
    \label{tab:my_table1}
    \end{table}
    \begin{table}[h]
    \centering
    \renewcommand{\arraystretch}{1.3}
    \title{\textbf{LPDA Element Response H-Plane}}\\
    \vspace{4 mm}
    \resizebox{\columnwidth}{!}{%
    \begin{tabular}{|c|c|c|c|c|c|}
    \hline
    \multirow{2}{*}{\textbf{Parameter(Units)}} &\multicolumn{5}{c|}{\textbf{Frequency in MHz}}
    \\
    \hhline{~-----}
    & \textbf{150} & \textbf{200} & \textbf{250} & \textbf{300} & \textbf{350 }\\
         \hline 
         Main Lobe(dBi) & 12.8 &12.6 & 12.9& 14.0 & 13.1\\
        Side Lobe(dB)& -19.2 &-18.0&-18.1& -2.2& -2.0\\
        Half power beam width & 84.5 & 117.4&147.4 & 57.6&88.5 \\
        \hline
    \end{tabular}}
    \caption{ The H-plane main-lobe, side-lobe and half-power beam widths obtained for an element (constituting 4 LPDAs) placed over a ground plane.}   \label{tab:my_table4}
    \end{table}


     \begin{table}[h]
    \centering
    \renewcommand{\arraystretch}{1.3}
    \title{\textbf{LPDA Element Response E-Plane}}\\
    \vspace{4 mm}
    \resizebox{\columnwidth}{!}{%
    \begin{tabular}{|c|c|c|c|c|c|}
    \hline
    \multirow{2}{*}{\textbf{Parameter(Units)}} &\multicolumn{5}{c|}{\textbf{Frequency in MHz}}
    \\
    \hhline{~-----}
    & \textbf{150} & \textbf{200} & \textbf{250} & \textbf{300} & \textbf{350 }\\
         \hline 
         Main Lobe(dBi) & 12.8 &12.5 & 10.8& 14.0 & 12.0\\
        Side Lobe(dB)& -17.0 &-13.4&-11.4& -11.8& -7.4\\
        Half power beam width & 22.5 & 17.7&14.4 & 11.6&10.5 \\
        \hline
    \end{tabular}}
    \caption{
    The E-plane main-lobe, side-lobe and half-power beam widths obtained for an element (constituting 4 LPDAs) placed over a ground plane.
 }   \label{tab:my_table3}
    \end{table}

\begin{figure}
    \centering
    \includegraphics[width=\linewidth]{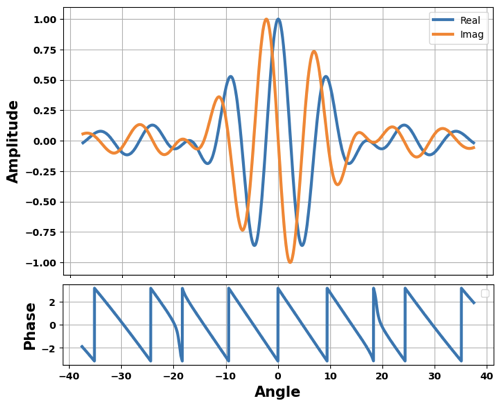}
    \caption{Interferometer simulation using a model built in MATLAB.  Fringe and phase plots of a transiting source for a time period of $\pm$ 2.5 hours from Zenith}
    \label{fig:subfig1}
\end{figure}

\subsection{Interferometer Response}

The expected response of the interferometer was first studied using a model developed in MATLAB. The MATLAB model correlated a simulated sky signal after incorporating the delay associated with an East-West interferometer baseline for a source that transits from the East to the West. The resulting correlation pattern, as shown in Figure \ref{fig:subfig1}, has peak responses when the antennas receive signals with no phase offset, and the correlated power decreases as delay is experienced in the signal path. The bottom plot shows the corresponding correlation phase as a function of the source's zenith angle.

\subsection{Characteristics of Front-end Receiver }

The low-noise amplifiers (LNAs), the 4-way RF combiner, the high-gain tubular receiver module, and the RFoF transceivers were characterised in laboratory environments before they were commissioned at the observatory. The tests were carriedout over a wider frequency range using a set of tones and broadband noise sources to study their operation over the 185 and 350 MHz and to verify the effectiveness of FM-band (88–108 MHz) rejection in the LNA filters. The band response obtained for the test input from the front-end electronics, including the LNA and tubular receiver, is shown in Figure  \ref{fig:BLOCK_DIAG} g.   The signal chain provided a total gain of 66 dB.

\section{Observational Results}

\begin{figure}[h]
    \centering
    \includegraphics[width=1\linewidth]{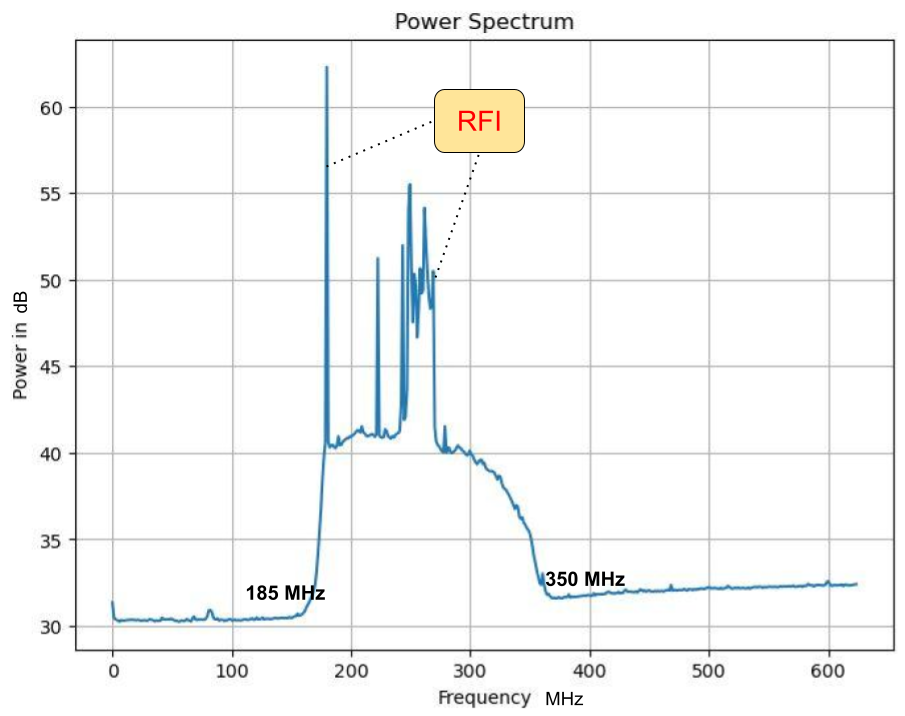}
    \caption{An element band response showing a few strong interfering signals observed at 185 MHz and between 240 and 270 MHz.   }
    \label{fig:powerspec}
\end{figure}

As part of the commissioning, we conducted observations using the two-element interferometer to assess RFI conditions before proceeding with observations. Figure \ref{fig:powerspec} shows a spectrum plot, with the interference signals seen between 185 and 350 MHz. Features seen in the 240-270 MHz range are from the satellites. In addition to these strong RFI signals, we observed occasional spurious signals at different frequencies throughout the day. 

\subsection{Observations of solar transits}

\begin{figure}[ht]
    \centering
    \includegraphics[width=\linewidth]{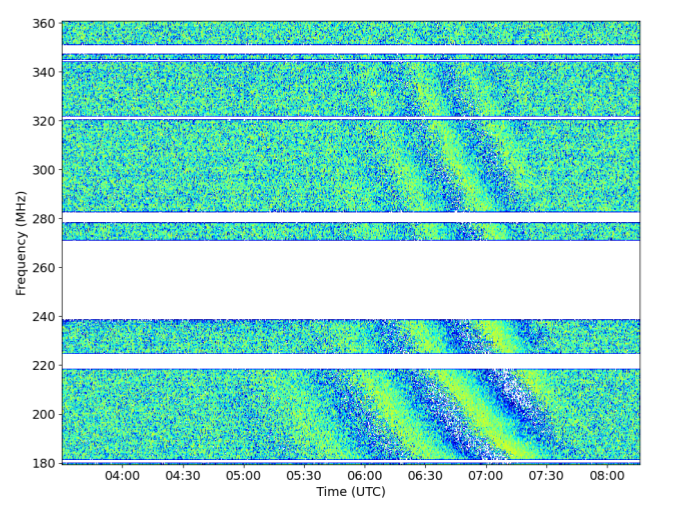}
    \caption{ The fringe drift occurred due to solar transit. The white patches seen are the masked RFI regions.}
    \label{fig:fringepattern}
\end{figure}

The observation was scheduled for seven hours during the day. The Sun transits over the two-element interferometer beam in approximately 3 hours. Figure \ref{fig:fringepattern} shows the fringe pattern obtained across the dynamic spectrum during a Solar transit. The intensity of the fringes reflects the varying phase coherence over time at different frequencies. 

\begin{figure}[h!]
    \centering
    \includegraphics[width=0.9\linewidth]{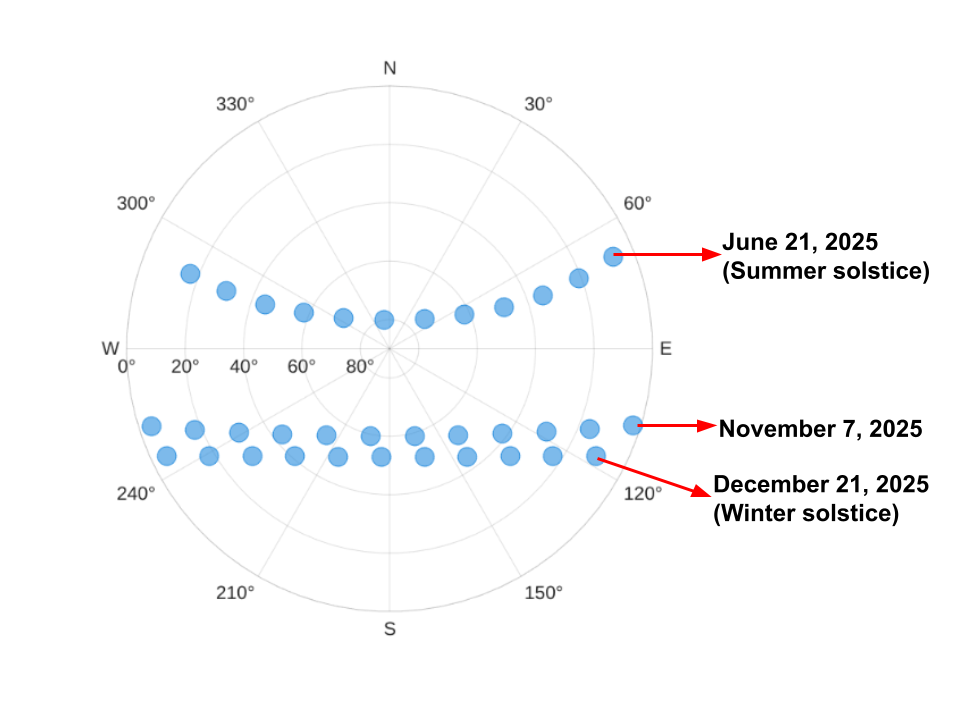}
    \caption{The cartoon depicts the solar path over TWIN on the Nov 7th 2025 flare. \citep{sunpath}.}
    \label{fig:sunpath}
\end{figure}

\begin{figure*}[ht!]
    \centering
    \includegraphics[width=0.95\linewidth]{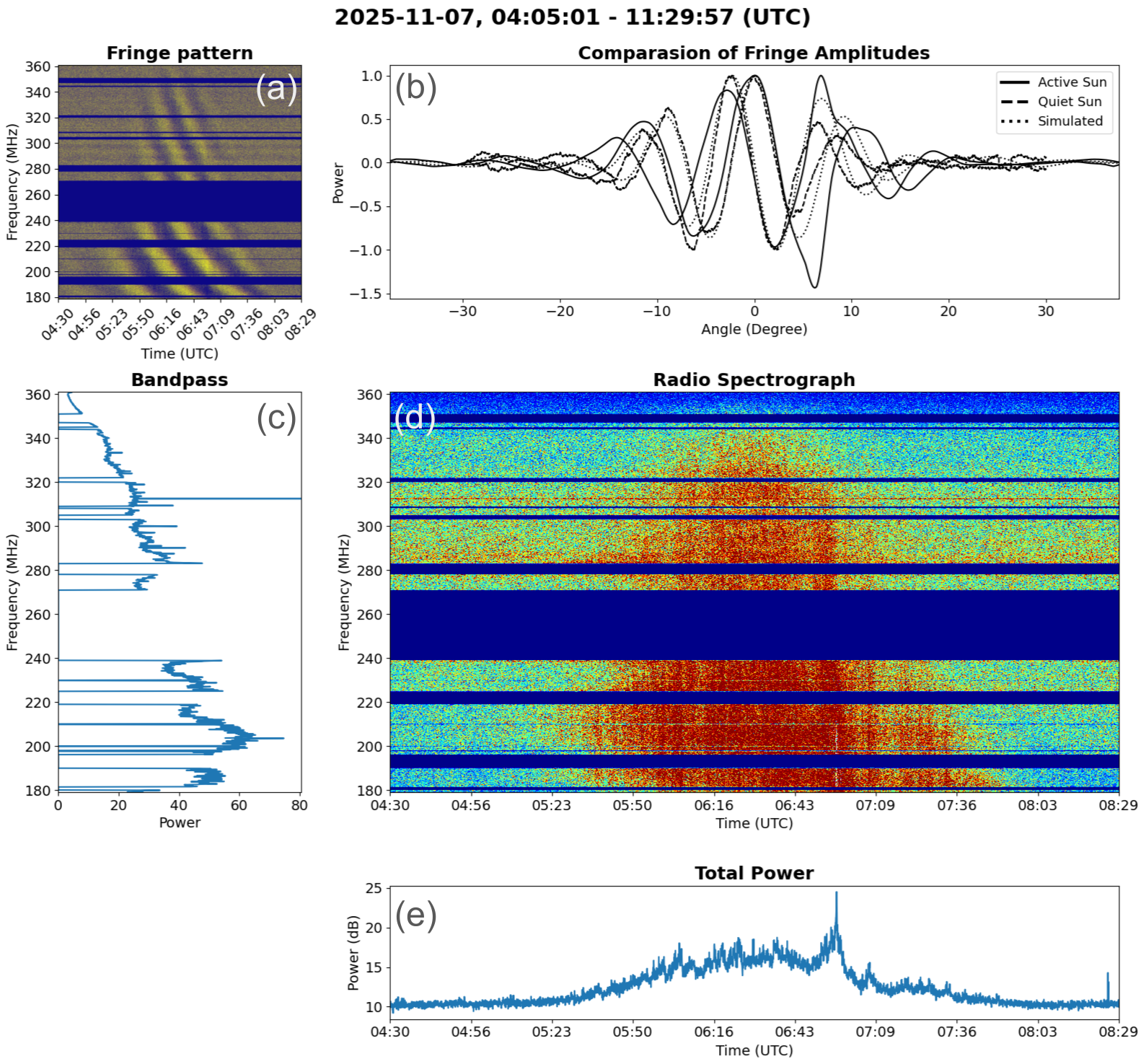}
    \caption{ November 7th 2025: A major solar flare detected by the TWIN. (a) Fringe drifting. (b) Comparison of fringe amplitude of Simulated signal, quiet Sun, and active Sun at 200 MHz. (c) Averaged bandpass. (d) Spectrograph of the A Type IV Solar Radio Burst observed on 11 November 2025 \citep{nov7} (e) Power of the received signals}
    \label{fig:07_11_25_SRB}
\end{figure*}

In such Solar observations, events such as solar flares and Coronal Mass Ejections (CMEs) can be detected. Solar flares are complex, dynamic events that release a large amount of energy as electromagnetic radiation and accelerated high-energy particles \citep{SF1, SF2}. Electromagnetic emissions are the earliest signatures of solar eruptions \citep{SF3}.  The five major types of radio emissions (also known as Solar Radio Bursts) are characterised by properties such as frequency bandwidth, drift rate, and duration, and are classified as Types I, II, III, IV, and V \citep{Graham}. The Two-Element interferometer can detect Type II, Type III, Type IV, and Type V bursts associated with solar flares and CMEs, likely visible in the 180 MHz to 350 MHz bandwidth.

\subsection{Data analysis and results}

The solar radio bursts appear as a drifting signature in time-versus-frequency plots. Among these signals, locally generated radio frequency interference (RFI) and satellite signals interfere with our band.  During analysis, it is necessary to mitigate RFI in the spectrum so that solar event signals can be recognised. A statistical analysis of the data helps identify and isolate the RFI-affected data. Manual inspection of the data can be cumbersome; hence, techniques such as the MAD-based algorithms are often considered. 

The TWIN observation scheduler \citep{likhit2025innovativewebtoolremote} provides a minimal set of options to assess the quality of the observed data, and a more robust analysis tool is discussed in \citep{PaneendraMSs} and \citep{TWIN2}, which provides comprehensive RFI mitigation in both the time and frequency domains.  

Our observatory is in the Northern Hemisphere, and the Sun’s positions at the two extreme latitudes over the TWIN in December and June are shown in the cartoon in Figure \ref{fig:sunpath}. On November 7th 2025, an intense solar flare occurred, and TWIN data captured this event. Figure \ref{fig:07_11_25_SRB} is the output after the analysis, vividly depicting the intense solar event \citep{nov7}. The event detection matches the e-CALLISTO Udaipur Solar Radio Spectrograph at the Udaipur Solar Observatory \citep{USO}.  Figure \ref{fig:sunpath} shows the solar path over TWIN on the Nov 7th 2025 flare.


\section{Future scope and discussion}

Manual detection of solar events can be difficult and time-consuming; therefore, we aim to automate it using Machine Learning techniques. A few challenges in this process include RFI detection and mitigation. RFI mitigation can also be automated using more modern techniques, involving AI and ML. A real-time RFI mitigation and solar burst detection pipeline is in development for inclusion in the pipeline. The data collected over the two years have been fully analysed, and the Solar flares detected are reported in our upcoming \citep{ursi1}.  In the present analysis, tracking delays have not been accounted for, thereby reducing the interferometric observation time to about 3 hours; we would incorporate tracking to enable longer-duration, sensitive observations. The entire Solar observation data set over the two-year period has been archived and is available in a data portal [\href{https://www.rri.res.in/obsastro}{https://www.rri.res.in/obsastro}].

\section{Conclusion}
We have designed and implemented a two-element interferometer using LPDA at the RRI Gauribidanur Radio Observatory. The data have been successfully acquired and processed in order to observe solar transits. Observing between 180 and 350 MHz helps to probe the Solar emissions from the upper Corona \citep{SF1}.  Nearly continuous solar monitoring has been carried out since July 2023, and several solar events have been detected in the observed data, as reported in \citep{TWIN2}. A simple-to-construct instrument such as TWIN would help with continuous observation and monitoring of events to study space weather and provide opportunities for student training in both instrumentation building and astronomical data analysis. 

\section{Acknowledgment}
We acknowledge the help received from the EEG members, RRI and our observatory staff, and their contributions to building the interferometer and making it operational. We sincerely thank S Somashekar for the guidance in the RFoF module design. We thank the workshop team, especially Srinath, Janardhana, and the entire Observatory staff, for helping us commission the system and for their involvement in its regular maintenance.


\bibliography{bibliography}
\end{document}